\documentstyle[11pt,epsf]{article}
\newcommand{\ac}{\acute}
\newcommand{\al}{\alpha}
\newcommand{\bt}{\beta}
\newcommand{\bv}{\begin{verbatim}}
\newcommand{\de}{\delta}
\newcommand{\ep}{\epsilon}
\newcommand{\ev}{\end{verbatim}}
\newcommand{\fr}{\frac}
\newcommand{\Ga}{\Gamma}
\newcommand{\ga}{\gamma}

\newcommand{\ka}{\kappa}
\newcommand{\La}{\Lambda}

\newcommand{\na}{\nabla}
\newcommand{\Om}{\Omega}

\newcommand{\Si}{\Sigma}
\newcommand{\si}{\sigma}

\newcommand{\td}{\tilde}

\newcommand{\th}{\theta}
\newcommand{\vb}{\verb}

\newcommand{\we}{\approx}
\newcommand{\be}{\begin{equation}}
\newcommand{\ee}{\end{equation}} 
\newcommand{\eei}{\end{equation}\indent\indent}
\newcommand{\bc}{\begin{center}}
\newcommand{\ec}{\end{center}}
\newcommand{\ber}{\begin{eqnarray}}
\newcommand{\ear}{\end{eqnarray}}
\newcommand{\ba}{\begin{array}}
\newcommand{\ea}{\end{array}}

\newcommand{\p}{\partial}

\def\case#1/#2{\textstyle\frac{#1}{#2} }
\begin{document}
\title{Non Metric Mass.}
\author{Mark D. Roberts, \\\\
Department of Mathematics and Applied Mathematics, \\ 
University of Cape Town,\\
Rondbosch 7701,\\
South Africa\\\\
roberts@gmunu.mth.uct.ac.za} 
\maketitle
\vspace{0.5truein}
\bc Eprint:  http://xxx.lanl.gov/abs/gr-qc/9812091 \ec
\bc Comments:  37 pages, no diagrams,  Latex2e.\ec
\bc Reference style:
in order appearing in text,\ec 
\bc with titles and MR (Mathematical Review) number where known.\ec
\bc 6 KEYWORDS:\ec
\bc Non-metricity,Schouten Geometry,Gravitional Waves,
Gravitational Potential,Dimesional Reduction,Cosmological Constant.\ec
\bc 1999 PACS Classification Scheme:\ec
\bc http://publish.aps.org/eprint/gateway/pacslist \ec
\bc 04.20CV,04.30+x,04.50+h,98.90+s,\ec
\bc 1991 Mathematics Subject Classification:\ec
\bc http://www.ams.org/msc \ec
\bc 83D05,83C35,83F05,83C55.\ec
\newpage
\begin{abstract}
Mass terms are often introduced into wave equations:  
for example introducing a 
mass term for a scalar field gives the Klein-Gordon equation
$(\Box^{2}-m^{2})\phi=0$.
Proceeding similarly with the metric of general relativity 
one recovers a vanishing mass term because $g_{ab;c}=0$.   
For non-metric theories $g_{ab;c}=-Q_{cab}$,  
so that the wave equation associated with 
the metric $(\Box ^{2}-M)g_{ab}=0$ no longer entails vanishing mass.   
This equation can be rewritten in the form
\bc $M(x)+\td{\na}_{a}Q_{.}^{a}+(\ep+\fr{d}{2}-2)Q_{a}Q_{.}^{a}=0$, \ec
where $\ep=0,  1,  2,  or~ 3$,  $d$ is the dimension of the spacetime,
and $Q$ is the object of non-metricity.
For any given non-metric theory 
it is possible to insert the metric into this wave equation 
and produce a non-metric mass.   
Alternatively one can choose this equation to be {\it a priori},  
and then try to construct theories for which it is the primary equation.   
This can be achieved using a simple Lagrangian theory.  
More ambitiously it is possible to investigate whether 
the introduction of non-metric mass has 
similar consequences to having a mass term in the 
Klein-Gordon and Proca equations:  
namely whether there are wave-like solutions, 
and what the rate of decay of the fields are.   
In order to find out a more intricate theory than the simple theory is needed.
Such a theory can be found by conformally rescaling the metric 
and then arranging that the conformal parameter 
cancels out the object of non-metricity in the Schouten connection.   
Once this has been achieved one can conformally rescale 
general relativity and then compare the properties of the wave equations.   
On the whole its consequences are similar to $m^{2}$ in the Klein-Gordon 
equation,  the main difference being $M$ is position dependent.   
The Proca $m^{2}$ breaks gauge invairiance,  
nothing similar happens for the non-metric mass $M$ 
or for the Klein-Gordon $m^{2}$.
The dynamics of the rescaled theory are not clearly defined;  
the best definition criterion is the 
initial value problem and this is taken to signify well-defined dynamics. 
\end{abstract}
\newpage
\section{Introduction}
\label{sec:intro}
In field theory operators such as $(\Box^{2}-m^{2})=(\na_{a}\na^{a}-m^{2})$ 
can be introduced to descibe massive fields.
General relativity has metric connection $\tilde{\na}_{a}g_{bc}=0$,  
and this forces $m$ 
to be zero if an operator of the above type is applied.   
The principle question addressed here 
is how to change the above so that non-zero mass operators 
can be applied to the metric.   
Changing the underlying geometry to a Schouten (1954) \cite{bi:schouten}
geometry there is a non-metric connection $\acute{\na}_{a}g_{bc}= -Q_{abc}$.
$\acute{\Box} g_{ab}$ is ambiguously defined 
(see \ref{eq:A0},  \ref{eq:B1},  \ref{eq:C2},  and \ref{eq:D3}),  
but it is not zero 
unless $Q_{abc}=0$ resulting in a non-metric mass related to $Q_{abc}$.   
For any given non-metric 
theory a non-metric mass can be calculated by inserting 
$Q_{abc}$ into $\acute{\na}_{c}g_{ab}$ 
and then using $M g_{ab}=\acute{\Box}^{2}g_{ab}$.   
Here however $(\acute{\Box}^{2}-M)g_{ab}=0$ is thought of 
as the {\it primary} equation and this 
leads to different theories than those considered hitherto.   
Previously Roberts (1986) \cite{bi:mdr86b} such 
theories have been referred to as massive theories,  
however in section \ref{sec:fpeq} it is shown that 
the resulting theories have nothing in common with 
linearized massive theories of gravitation 
as described by the Fierz-Pauli \cite{bi:FP} equation.   
In section \ref{sec:simple} a simple non-metric theory,  
determined by Hilbert's Lagrangian and a Lagrangian 
of similar form to the Klein-Gordon 
scalar Lagrangian but with the object of non-metricity 
replacing the scalar field is presented.   
Physically more interesting is a theory introduced in section 
\ref{sec:rescaled} which uses the rescaling property of non-metric spaces.   
This theory has problems with the consistency of the dynamics.   
Consistent dynamics are taken to be those for which there is a well posed 
intial value problem and this is discussed in section \ref{sec:ivp}.   
Previous examples of the application of non-metricity in relativity are 
the original theory of Weyl (1918) \cite{bi:weyl},  
scale covariant theories Canuto et al (1972) \cite{bi:CAHT},  
and quantum theories Smolin (1975) \cite{bi:smolin}.
Non-metric theories have also been discused in
Coley (1983) \cite{bi:coley},
Hochberg and Plumien (1991) \cite{bi:HP},
Blanchet (1992) \cite{bi:blanchet},
Poberii (1994a) \cite{bi:poberiia},
Hehl et al (1995) \cite{bi:HMMN},
Aldrovandi et al (1998) \cite{bi:ABCP},
Socorro et al (1998) \cite{bi:SLMM},
and Chen and Nester (1998) \cite{bi:CN}.
Conformal invariance suggests that non-metricity might be related to 
2-dimensional dilaton theories,  
see for example Banks and O'Loughlin (1991) \cite{bi:BOL}.   
Non-metric theories are different from 2-metric theories,
Aicheburg and Mansouri (1972) \cite{bi:AM},
as the underlying geomerty is different.

Having introduced theories with non-metric mass operators,  
the next question to be addressed is {\it why}.   
The main property of operators such as $(\Box^{2}-m^{2})$ is that they lead to 
differential equations which have solutions which are no longer null 
(or light-like).   
One reason is that it might be possible to explain faster-than-light effects.
Observationally apparent faster than light extra-galatic radio sources 
have been known for some time,  Cohen et al (1977) \cite{bi:cohen}.   
Now there are also observations, Mirabel and Rodriguez (1994) \cite{bi:MR},
that suggest that there are super-luminal sources in the galaxy.   
These observations are further discussed in 
Pearson and Zenus (1987) \cite{bi:PZ}.
The speed of propagation of the gravitational field might be 
connected to faster-than-light effects.
It has been suggested that the speed of propagation of the gravitational
field be measured directly,  Roberts (1987a) \cite{bi:mdr87a}. 
If one looks at relativistic gravitational theories there 
are at least four ways to approach whether 
gravitational interaction is null or not.  
The {\it first} is to look at the shock waves (or charachteristics)  
Synge (1960) \cite{bi:synge}.   
This is not a good indicator as for example the shock waves of 
the Klein-Gordon equation are null.   
The {\it second} is to look at weak field approximations.  
It is shown in section \ref{sec:ds} that in 
DeSitter (1917) spacetime \cite{bi:DS} these 
lead to the Fierz-Pauli equation with mass $m^{2}=\fr{2}{3}\La$,
so that again these do not suggest lightlike propagation.   
The {\it third} is to look at exact solutions.   
It has been shown that in Quadratic Lagrangian theory that wave 
solutions are not necessarily null,  Roberts (1994) \cite{bi:mdr94}.   
Once one has a candidate exact non-null solution the speed 
of energy transfer can be tested.   
In gravitational theories the total energy is only 
unambiguous for asymptotically flat spacetime,  
for other spacetimes there are 
several ways of approaching this,  see for example
Gustavo (1994) \cite{bi:gustavo},
and Chen and Nester (1998) \cite{bi:CN}.
The {\it fourth} way of investigating non-null propagation 
is to consider non-metric operators and this is done in section \ref{sec:pot}.

The other important property of operators such as $(\Box^{2}-m^{2})$ 
is that they alter the rate of decay of the fields.   
The best known example of this is the Yukawa potential,  which shows 
that the Klein-Gordon field decays quicker with a positive mass term.   
There are solutions to the scalar-Einstein equations,  Roberts (1996)
\cite{bi:mdr96};  
however there are none to the massive scalar-Einstein 
equations which suitably generalize the Yukawa potential.   
A different potential might have bearing on large scale dynamics:  
from the scale of the outer solar system Roberts (1987b) \cite{bi:mdr87b} 
to the whole Universe dynamical discrepencies grow.
For example constant galactic rotation curves Rubin (1983) \cite{bi:rubin},
are incompatible with dynamics from the visible mass of galactic stars,
and the discrepency is even greater for clusters of galaxies.
The longer the lenght scale the greater the discrepency.   
Potentials from Quadratic Lagrangian theory have 
been used to attempt to explain constant galactic rotation curves,
Schmidt (1990) \cite{bi:schmidt} and Roberts (1991b) \cite{bi:mdr91b}.   
How non-metric mass effects potentials is discussed in section \ref{sec:pot}.
The rate of decay for weak massive gravitation has been discused in
Freund et al (1969) \cite{bi:FMS},
Datta and Ranna (1979) \cite{bi:DR},
and Ford and Van Dam (1980) \cite{bi:FVD}

Section \ref{sec:ss} shows that the non-metric wave equation 
on its own can have an 
arbitrary mass parameter while being within observable bounds.   
In section \ref{sec:cc} the extent to which it is possible to reinterpret the 
anomolously high theoretical value of the cosmological constant as 
simply a manifestation of non-metric mass is discussed.   
To do this it is necessary to generalize 
Freund-Rubin (1980) \cite{bi:FR} compactification,
by relaxing the usual restrictions put on the coordinates,
this is similar in style to some of the 5-dimensional approaches 
discussed in Overduin and Wesson (1997) \cite{bi:OW}.
The cosmological implications of the rescaled 
theory are briefly mentioned in section \ref{sec:cos}.

The conventions used are: signature-+++,  $d$ is the number of dimensions.   
Square, round,  and Schouten \cite{bi:schouten} brackets are defined by
\be
2T_{[ab]}=T_{ab}-T_{ba},
\label{eq:sq}
\ee
\be
2T_{(ab)}=T_{ab}+T_{ba},
\label{eq:rd}
\ee
\be
y_{\vb+{+abc\vb+}+}=y_{abc}+y_{cba}-y_{bac},
\label{eq:sc}
\ee
respectively.   
The Riemann tensor,  Ricci tensor,  Ricci scalar,  and Einstein tensor are 
defined by
\be
R^{a}_{bcd}=2\p_{[c}\Ga^{a}_{.d]b}+2\Ga^{a}_{f[c}\Ga^{f}_{.d]b},
\label{eq:rie}
\ee
\be 
R_{bd}=R^{a}_{bad},
\label{eq:ric}
\ee
\be
R=R^{a}_{a},
\label{eq:rs}
\ee
\be
G_{ab}=R_{ab}-\fr{1}{2}g_{ab}R,
\label{eq:ein}
\ee
respectively.   
The connection use is shown by a mark above geometric objects: 
thus $\acute{R}^{a}_{.bcd}$ is the Riemann 
tensor constructed with connection $\Ga$ and $\tilde{R}^{a}_{.bcd}$ 
is the Riemann tensor constructed with Christoffel connection
\be 
\vb+{+^{a}_{bc}\vb+}+=\fr{1}{2}g^{ad}g_{\vb+{+dc,b\vb+}+};
\label{eq:Chrcon}
\ee   
similarly $\bar{\Box}^{2}=\p_{a}\p^{a}$ 
and $\tilde{\Box}^{2}=\na_{a}\na^{a}$,  
where $\na^{a}$ is 
covariant derivative using the Christoffel connection. 
$\acute{\Box}^{2}$ is ambiguous 
and will be defined in section \ref{sec:nmme}.
The commutation of covariant derivatives when the torsion vanishes is 
\be
X_{ab;cd}-X_{ab;dc}=X_{ae}R^{e}_{.bde}+X_{eb}R^{e}_{.acb}.
\label{eq:comdev}
\ee

Stresses used include the Klein-Gordon scalar stress
\be
T_{ab}=2\psi_{a}\psi_{b}-g_{ab}\left(\psi_{c}\psi^{c}+V(\psi^{2})\right),
\label{eq:kgstress}
\ee
where the potential $V(\psi^{2})$ usually is taken to have 
soley the mass term $m^{2}\psi^{2}$,  and the fluid 
stress,  c.f. Ellis \cite{bi:ellis} p.116
\be
T_{ab}=(\mu+p)U_{a}U_{b}+q_{a}U_{b}+U_{a}q_{b}+p~ g_{ab}+\Pi_{ab},
\label{eq:fluidstress}
\ee
where $q_{a}U^{a}_{.}=0,  \Pi^{a}_{.a}=0,  \Pi_{ab}U^{a}_{.}=0$.   
$U_{a}$ is a timelike vector field with $U_{a}U^{a}_{.}=-1$,  
$\mu$ is the energy density of matter measured by $U_{a}$,  
$q_{a}$ is the energy flux relative to $U_{a}$,  
$p$ is the isotropic pressure,  
$\Pi_{ab}$ is the anistropic pressure.   
Einstein's field equations are
\be
G_{ab}=\ka T_{ab},
\label{eq:efldeq}
\ee
where $\ka=\fr{8pG}{c^{2}}$;  
this constant is often absorbed into the fluid or scalar field.
\section{The Non-metric Mass Equation.}
\label{sec:nmme}
In field theory it is possible to produce massive fields such as the
massive scalar field,  which obeys the Klein-Gordon equation
\be
(\Box^{2}-m^{2})\phi=0,
\label{eq:KG}
\ee
and the massive vector field,  which obeys Proca's equation
\be           
(\Box^{2}-m^{2})A_{a}=0.
\label{eq:Pr}
\ee
For weak massive spin 2 fields there is the Fierz-Pauli (1939) 
equation \cite{bi:FP}
\be
(\Box^{2}-m^{2})h_{ab}=0.
\label{eq:FPeq}
\ee
If the same type of equation is introduced for the metric
\be
(\Box^{2}-M)g_{ab}=0,
\label{eq:An}
\ee
then in general relativity $M$ must be zero because
\be
\tilde{\na}_{c}g_{ab}=0.
\label{eq:fgr}
\ee
To have a theory with non-vanishing mass in equation \ref{eq:An} it is 
necessary to generalize the under-lying geometry so that there is a 
non-vanishing object of non-metricity
\be
\acute{\na}_{a}g_{..}^{bc}\equiv Q_{a..}^{bc},
\label{eq:Qdef}
\ee
Non-metric mass theory is concerned with the consequences of \ref{eq:An}
when the covariant derivatives involve the object of non-metricity 
so that the mass is non-vanishing,  the resulting equation is called the 
non-metric mass equation.   
For the Klein-Gordon equation \ref{eq:KG} 
$m^{2}$ is identified as the mass because:
i)the equation is a consequence of applying the quantization rules to
a point particle,  ii)the rate of decay of the field $\phi$  is faster with
a non-vanishing mass,  iii)the wave solutions are slower-than-light for
$m$ positive and real;  and the last two of these are investigated for
non-metric mass in section \ref{sec:pot}.

Using
\be
g_{..}^{ab}g_{bc}=\de^{a}_{c},
\label{eq:Kd}
\ee
and \ref{eq:Qdef} gives
\be
\acute{\na}_{a}g_{bc}=-Q_{abc}.
\label{eq:Qsub}
\ee
In a Schouten (1954) \cite{bi:schouten} geometry the connection is
\be
\Ga^{a}_{bc}=\vb+{+^{a}_{bc}\vb+}++\fr{1}{2}N^{a}_{.bc},
\label{eq:Ga}
\ee
where $\vb+{+^{a}_{bc}\vb+}+$ is the Christoffel connection
and $N^{a}_{.bc}$ is given by
\be
N^{a}_{.bc}=g_{..}^{cd}(-2S_{\vb+{+bdc\vb+}+}+Q_{\vb+{+bdc\vb+}+}),
\label{eq:Schcon}
\ee
where the Schouten bracket given by equation \ref{eq:sc} is used.   
Here the torsion $S_{bdc}$ is taken
to vanish and the object of non-metricity and connection are restricted 
to the form
\ber
\acute{\na}_{a}g_{..}^{bc}\equiv +Q_{a..}^{bc}=+Q_{a}g_{..}^{bc},\nonumber\\
\acute{\na}_{a}g_{bc}=-Q_{abc}=-Q_{a}g_{bc}.
\label{eq:deQ}
\ear
Equations \ref{eq:deQ} reduce \ref{eq:Schcon} to the connection
\be
N^{a}_{.bc}=(Q_{b}\de^{a}_{c}+Q_{c}\de^{a}_{b}-g_{bd}Q^{a}_{.}),
\label{eq:Qnm}
\ee
of a semi-metric geometry (also called a Weyl geometry).   Using these 
connections tensors can be defined as for the equations \ref{eq:rie}
\ref{eq:ric},  \ref{eq:rs},  and \ref{eq:ein};
also the tensor
\be
P^{a}_{.bcd}=\acute{R}^{a}_{.bcd}-\tilde{R}^{a}_{.bcd},
\label{eq:prdiff}
\ee
is of interest.   Using \ref{eq:rie} and \ref{eq:Qnm} 
equation \ref{eq:prdiff} becomes
\ber
P^{a}_{.bcd}=\fr{1}{2}\left((Q_{dc}-Q_{cd})\de^{a}_{b}
                  +Q_{bc}\de^{a}_{d}-Q_{bd}\de^{a}_{c}
                  -Q^{a}_{.c}g_{bd}+Q^{a}_{d}g_{.cb}\right)\nonumber\\
            +\fr{1}{4}\left(-Q_{b}Q_{c}\de^{a}_{d}+Q_{b}Q_{d}\de^{a}_{d}    
                  -g_{cb}Q^{a}_{.}Q_{d}+g_{bd}Q^{a}_{.}Q)  
                  +(g_{bc}\de^{a}_{d}-g_{bd}\de^{a}_{c})Q_{e}Q^{e}_{.}\right).
\label{eq:pie}
\ear
Contracting
\ber
P_{bd}=P^{a}_{.bad}=\fr{1}{2}\left(Q_{db}+(1-d)Q_{bd}\right)
                   -\fr{1}{2}g_{bd}Q^{a}_{.a}\nonumber\\
                   +\fr{d-2}{4}\left(Q_{b}Q_{d}-g_{bd}Q_{a}Q^{a}_{.}\right),
\label{eq:pic}
\ear
this tensor is not symmetric unless $Q_{a}$ 
is a gradient vector ($Q_{a}=Q_{,a}$).
$P_{ab}$ occurs in non-metric field equations,  
if $Q_{a}$ is not a gradient vector it results in an asymmetric stress.

Another place where a different connection is used is in the weak field
perturbations off a background field.
The metric is taken to be of the form
\be
g_{ab}=\bar{g}_{ab}+h_{ab},~~~g^{ab}=\bar{g}^{ab}-h^{ab},
\label{eq:barm}
\ee
where $\bar{g}_{ab}$ is a given background field metric and $h_{ab}$ is a 
small perturbation of the metric.   Similarly to \ref{eq:Ga} 
decompose the connection and take $N^{a}_{.bc}$ to be given by $H^{a}_{.bc}$
where
\be
H_{abc}=h_{ba;c}+h_{ca;b}-h_{bc;a}.
\label{eq:Hh}
\ee
Let ``;'' be the covariant derivative using the Christoffel symbol of the 
background field metric.   For any connection which is a sum of the
Christoffel symbol and a tensor $H^{a}_{.bc}$,  the Riemann tensor is
\be
R^{a}_{.bcd}=\bar{R}^{a}_{.bcd}+H^{a}_{.[d|b|c]}+H^{a}_{.eb}S^{e}_{cd}
                               +\fr{1}{2}H^{a}_{.[c|e|}H^{e.b}_{.d]}.
\label{eq:Ra}
\ee
In the present case cross terms in $H_{abc}$ and the torsion $S^{a}_{.bc}$
are taken to vanish.   Substituting equation \ref{eq:Hh} 
into equation \ref{eq:Ra} and using \ref{eq:comdev} for the commutation of
covariant derivatives the Riemann tensor becomes
\be
R^{a}_{.bcd}=\bar{R}^{a}_{.bcd}-\fr{1}{2}h^{a}_{e}\bar{R}^{a}_{.bcd}
                               -\fr{1}{2}h_{be}\bar{R}^{ea}_{..cd}
+\fr{1}{2}(h^{.c}_{d.;bc}-h^{..a}_{db.c}-h^{.a}_{c.;bd}+h^{..a}_{cb;.d})
\label{eq:Rb}
\ee 
contracting and again using the commutation of 
covariant derivatives \ref{eq:comdev}
\be
R_{bd}=\bar{R}_{bd}-\fr{1}{2}h^{fe}\bar{R}_{ebcd}
       +\fr{1}{2}h_{be}\bar{R}^{e}_{.d}+\fr{1}{2}h_{de}\bar{R}^{e}_{.b}
+\fr{1}{2}\left(h^{.c}_{d.;cb}+h^{.c}_{b.;cd}-\Box^{2}h_{db}-h_{;bd}\right),
\label{eq:Rc}
\ee
where $h=h^{c}_{.c}$.   The tensor $h_{ab}$ has ten components with four
degrees of freedom.   The Plebanski-Ryten (1961) \cite{bi:PR} 
gauge condition is
\be
[(-g)^{w}_{.}g^{ab}_{..}]_{,b}=0,
\label{eq:PlRy}
\ee
with $w$ arbitrary.   When $w=1$ the Ricci scalar's 
second derivative terms vanish.
For weak fields this coordinate condition gives the gauge condition
\be
h_{a.,b}^{b}=w~h_{,a},
\label{eq:whab}
\ee     
when $w=\fr{1}{2}$ this is the harmonic gauge condition.   
The four degrees of freedom in $h_{ab}$ can be removed by applying 
the harmonic gauge condition.   When this is done equation \ref{eq:Rc} becomes
\be
R_{bd}=\bar{R}_{bd}-h^{ef}\bar{R}_{ebfd}+\fr{1}{2}h_{fe}\bar{R}^{e}_{.d}
       -\fr{1}{2}\Box^{2}h_{bd}.
\label{eq:Rd}
\ee

Returning to non-metric spaces consider their conformal properties.
Conformally rescaling the metric from $g_{ab}$   
(usually a metric of a space with Christoffel connection) 
to $\hat{g}_{ab}$ a Schouten space i.e.
\be
\hat{g}_{ab}=\Om(x)^{2}g_{ab},
\label{eq:crs}
\ee
and then using \ref{eq:Qsub} and \ref{eq:Schcon} gives
\ber
\acute{\na}_{c}\hat{g} _{ab}
&=&(-\Om^{2}Q_{c}+\p_{c}(\Om^{2}))g_{ab}\nonumber\\   
&=&(-Q_{c}+\p_{c}ln(\Om^{2}))\hat{g}_{ab}\nonumber\\
&=&-\hat{Q}_{cab}\nonumber\\
&=&-\hat{Q}\hat{g}_{ab},
\label{eq:215}
\ear
hence conformally rescaling the metric,  as in \ref{eq:pic},  
and simultaneously transforming the object of non-metricity
\be
\hat{Q}_{a}=Q_{a}-\p_{a}\ln(\Om^{2}),
\label{eq:sim}
\ee
gives the same connection,  and hence the same Riemann tensor.   This can also
be seen directly by requiring that the connections 
\ref{eq:Chrcon} and \ref{eq:Schcon} cancel.   
The equivalence of connections is frequently used to relate the solutions of 
general relativity to corresponding non-metric solutions.  To see this note 
that if $g_{ab}$  is a solution to the field equations of general relativity 
(as given by equations \ref{eq:efldeq} where $Q_{a}=0$),  
then dropping the hat on the object of 
non-metricity in the Schouten space and conformally rescaling the metric,
as in \ref{eq:crs},  gives a solution to the field equations
\ber
\acute{G}_{ab}&=&\tilde{G}_{ab}+P_{ab}-\fr{1}{2}g_{ab}P\nonumber\\
              &=&\tilde{G}_{ab}+[Q_{ba} +(1-d)Q_{ab}] 
                    +\fr{d-2}{4}Q_{a}Q_{b}
      +g_{ab}\fr{d-2}{2}[Q_{c.}^{.c}+\fr{d-3}{4}Q_{c}Q_{.}^{c}]\nonumber\\
              &=&\ka T_{ab},
\label{eq:feq}
\ear
where the object of non-metricity is given by
\be
Q_{a}=-\p_{a}\ln(\Om^{2}).             
\label{eq:218}
\ee
This equation implies that the object of non-metricity 
$Q$ is a gradient vector,
therefore $P_{ab}$   is a symmetric tensor and as $\tilde{R}_{ab}$   
is a symmetric tensor,  
thus $T_{ab}$ remains a symmetric stress tensor after rescaling.   
Equation \ref{eq:218} can be integrated to give
\be
\Om^{2}=\exp(-Q),
\label{eq:expQ}
\ee
where the constant of integration can be absorbed into $Q$ as it is only 
derivatives of $Q$ that appear in geometrical objects.   
The minus sign in front of $Q$ is unfortunately fixed 
as can seen from \ref{eq:218}.
The rescaling property is illustrated for 
an exact solution \ref{eq:Wsol}.

In non-metric geometry a massive wave equation of the form can be 
defined in four ways:
\be
(g_{..}^{ab}\ac{\na}_{a}\ac{\na}_{b}-M)g_{cd}=0,
~~~~~~~~~~~~~~(for~ \ep=0),
\label{eq:A0}
\ee
\be
(\ac{\na}_{a}g_{..}^{ab}\ac{\na}_{b}-M)g_{cd}=0,
~~~~~~~~~~~~~~(for~ \ep=1),
\label{eq:B1}
\ee
\be               
(g_{..}^{ab}\ac{\na}_{a}\ac{\na}_{b}+M)g_{..}^{cd}=0,
~~~~~~~~~~~~~~(for~ \ep=2),
\label{eq:C2}
\ee
\be
(\ac{\na}_{a}g_{..}^{ab}\ac{\na}_{b}+M)g_{..}^{cd}=0,
~~~~~~~~~~~~~~(for~ \ep=3).
\label{eq:D3}
\ee
Substituting equations \ref{eq:Qsub} and \ref{eq:Schcon} 
into \ref{eq:A0} and \ref{eq:B1} gives
\be
Mg_{cd}=-g_{..}^{ab}\ac{\na}_{a}Q_{bcd}-\ep  Q_{a..}^{.ab}Q_{bcd},
\label{eq:AB}
\ee
and into \ref{eq:C2} and \ref{eq:D3} gives
\be
Mg_{..}^{cd}=-g_{..}^{ab}\ac{\na}_{a}Q_{b..}^{.cd}
              -(\ep-2)Q_{a..}^{ab}Q_{b..}^{.cd},
\label{eq:CD}
\ee
writing out the covariant $\Ga$ derivatives involves unwieldy equations;  
however for a semi-metric space \ref{eq:deQ}
all four equations can be collected together
in a straightforward equation,  here called the non-metric mass equation
\be
M(x)+\td{\na}_{a}Q_{.}^{a}+(\ep+\fr{d}{2}-2)Q_{a}Q_{.}^{a}=0,
\label{eq:nmass}
\ee
where $d$ is the dimension of the spacetime.   The mass $M(x)$ is a function 
of undetermined sign,  this permits the definitions \ref{eq:A0} 
and \ref{eq:B1} to have the 
opposite sign of \ref{eq:C2} and \ref{eq:D3},  
which in turn allows all four equations to 
be put in the form \ref{eq:nmass};  
this is why $M$ is used for non-metric mass 
\ref{eq:CD} whereas $m^{2}$  occurs in the Klein-Gordon equation \ref{eq:KG},
the Proca equation \ref{eq:Pr},  and the Fierz-Pauli equation \ref{eq:FPeq}.
Say that there is given a non-metric spacetime (which can originate from 
Weyl's or any other non-metric theory),  
then a non-metric mass can be calculated 
by substitution of $Q_{a}$ into \ref{eq:nmass}.   
Unlike the Barut and Haugen (1972) \cite{bi:BH} mass the
mass $M(x)$ is not conformally invariant.   This can be seen by rescaling the
metric of the non-metric spacetime and using \ref{eq:sim} 
to find the rescaled object of non-metricity;  
then using \ref{eq:nmass} again gives a rescaled mass which 
is not the same as the original mass.
\section{The Simple Theory.}
\label{sec:simple}
Equation \ref{eq:nmass} can generate a mass for any non-metric theory,  
but the theories discussed here are motivated by using this equation 
as a starting point.   The simplest theory can be found by noting that 
with the choice $\ep=0$,  $d=4$,  $Q_{a}$  is a gradient vector and
\be
M(x)=-M^{*}\fr{dV(Q)}{dQ}  
\label{eq:mstar}
\ee                                                                          
where M* is a constant,  the non-metric equation \ref{eq:nmass}
simplifies to
\be
\td{\Box}Q-M^{*}\fr{dV(Q)}{dQ}=0,
\label{eq:td}
\ee
and this equation can be derived from the Lagrangian
\be
L_{Q}=-\fr{1}{2}Q_{a}Q_{.}^{a}-\fr{1}{2}M^{*}V(Q).   
\label{eq:LQ}
\ee                           
A choice for a massive theory of gravity is to choose that the dynamics 
be given by the Lagrangian $L_{Q}$ and by Hilbert's Lagrangian $L_{H}$
\be
L_{H}=\sqrt{-g}\td{R},     
\label{eq:hl}
\ee
this choice is here called the simple non-metric massive theory of 
gravitation,   or the simple theory for short.   Such a theory is 
similar to Einstein-scalar theory,  but based on a different geometry.   
It has the advantage that it has a Lagrangian formulation:  resulting in 
its dynamics being simpler and more secure than would otherwise be the 
case.  The disadvantage is that the field equations for it are
\be
\td{G}_{ab}=Q_{a}Q_{b}-\fr{1}{2}g_{ab}\left(Q_{c}Q_{.}^{c}+M^{*}V(Q)\right)
            +\ka T_{ab},
\label{eq:tG}
\ee
and these do not have the rescaling properties of the field 
equations \ref{eq:feq};  similarly the rescaling properties do not 
hold if the left hand side of \ref{eq:mstar} is replaced by 
$\ac{G}_{ab}$.   
The rescaling property is useful as it allows solutions to the 
field equations \ref{eq:feq} to be found if solutions to the field 
equations of general relativity \ref{eq:efldeq} are known,  and this 
does not happen for the simple theory.
\section{The Rescaled Theory}
\label{sec:rescaled}
The theory which uses rescaling consists of the field equations 
\ref{eq:feq} and the non-metric mass equation 
\ref{eq:nmass},  with as yet no other equation.   
Rescaling \ref{eq:nmass} from a Schouten geometry 
(scripted S) to a metric in a space with only Christoffel connection 
(scripted with E) gives
\ber
0&=&M(x)+\td{\na}Q^{(S)a}+(\ep+\fr{d}{2}-2)Q^{(S)}_{a}Q^{(S)a}\nonumber\\ 
&=&M(x)+\exp(+Q)[\td{\na}^{(E)}_{a}Q_{.}^{a}+(\ep-1)Q_{a}Q_{.}^{(E)a}].
\label{eq:rsm}
\ear
The theory consisting of \ref{eq:feq} and \ref{eq:rsm},
in $d=4$ dimensions, 
is called the non-metric massive theory of gravitation rescaled 
from general relativity,  or the rescaled theory for short.   
The rescaled theory is physically more interesting than 
the simple theory,  but because it is not Lagrangian formulated 
there are problems with the consistency of its dynamics.   

A Lagrangian formulation is no guarantee of unique consistent 
field equations,   for example just varying the metric as
opposed to varying the metric and connection can give different 
results,  see Heyl and Kerlick (1978) \cite{bi:HK},
also for Quadratic Lagrangian theory see Buchdahl (1979) \cite{bi:buchdahl} 
and Shalid-Saless (1991) \cite{bi:SS}.   
Also there are dynamical theories for which no Lagrangian formulation exists.
For example the theory of motion of a charged particle 
see Roberts (1989a) \cite{bi:mdr89a}, 
where there are terms such as the Hobbs,  Dewitt-Brehme, 
and Lorentz-Dirac terms which cannot be derived from a Lagrangian.
The possibility of the equations of motion containing more information 
than the Lagrangian has also been discussed in 
Falconi (1994) \cite{bi:falconi}.

The rescaled theory with $\ep=1$ has been discussed in the authors 
thesis,  Roberts (1986) \cite{bi:mdr86b}.   
It is not unusual for theories involving 
the object of non-metricity to have a different number of variables 
than equations,  for example Einstein and Bergman's \cite{bi:bergman}
variant of Weyl's 
theory has 10 equations in 13 variables.   Here the rescaled theory 
consists of 11 geometrical objects,  10 components of the metric $g_{ab}$,  
and as the rescaling property requires the object of non-metricity $Q$ 
to be a gradient vector,  only one other geometric object $Q$;  it also has 
11 physical quantities,  rescaling forces the stress $T_{ab}$  
to be symmetric 
and therefore it has 10 components,  there is also the non-metric 
mass $M(x)$;  the dynamical equations suggested so far are the 10 field 
equations \ref{eq:feq}  
and the non-metric mass equation \ref{eq:nmass};  
thus viewed as a linear problem,  and barring degeneracy of the equations,  
the dynamics are consistent.   The dynamics are non-linear and in this
section the addition of more equations is discussed;  in section \ref{sec:ivp}
it is found that the most elegant formulation of the initial value 
problem includes one of these additional equations.

The field equations \ref{eq:feq}
and the non-metric equation \ref{eq:nmass} may not
fully determine the dynamics of the rescaled theory.   
In this section further 
equations are constructed and their value briefly assessed.

An additional equation is
\be
M g_{ab}=\al T_{ab},
\label{eq:add1}
\ee
where $\al$ is a constant.   This equation suffers from two drawbacks:   
{\it firstly} it
over constrains the system by giving 10 further equations,  
and {\it secondly} it 
gives bizarre results for spherically symmetric electromagnetic fields;  
because of these drawbacks this equation is not used.

Another additional equation is
\be
Q_{a}Q_{.}^{a}=\al T_{[ab]}T^{[ab]},
\label{eq:add2}
\ee
where $\al$ is a constant,   the   dependence of this equation equates 
the degree of asymmetry of the stress tensor to the object of non-metricity,
thus it vanishes in the rescaled theory,  but might be of use in other 
non-metric theories.

The most successful additional equation is the trace equation
\be
M=\al T,
\label{eq:mtr}
\ee
where $\al$ is a constant,  $\al T$ is taken
to be the trace of the stress before rescaling times a constant.  
This choice has the possibility of being derived from a quantum theory as a 
trace anomaly effect.   Let $W_{a}$  be a unit timelike vector field,  
so that $W_{a}W_{. }^{a}=-1$.   The weak energy condition and the timelike 
convergence condition are obeyed if
\be
T_{ab}W^{a}W^{b}\geq0,
\label{eq:wec}
\ee
\be
R_{ab}W^{a}W^{b}\geq0,
\label{eq:tlc}
\ee
respectively,  see for example Hawking and Ellis (1973) \cite{bi:HE} p.89,95.
Assuming the field equations \ref{eq:feq}
\be
\ac{G}_{ab}W^{a}W^{b}=T_{ab}W^{a}W^{b}\geq0,
\label{eq:acG}
\ee
the weak energy condition is obeyed.   Thus
\be
\ac{R}_{ab}W^{a}W^{b}+\fr{1}{2}\ac{R}\geq0.
\label{eq:acR}
\ee
Assuming the time-like convergence condition is to apply to the Ricci tensor
$\ac{R}$ not $\td{R}$,  and noting that $T=-\ac{R}$ gives
\be
T=\fr{M}{\al}\leq0,~~~~~~~~(for~~~ \al\neq0).
\label{eq:acT}
\ee
This implies that $M$ and $\al$ must be of opposite sign for the energy 
conditions to be obeyed.   For a perfect fluid  $T=3p-\mu$;   
therefore equation \ref{eq:A0} becomes
\be
\left(g_{..}^{ab}\ac{\na}_{a}\ac{\na}_{b}-(3p-\mu)\right)g_{ab}=0,
\label{eq:acg}
\ee
with similar equations for $B$, $C$, and $D$.   
Equation \ref{eq:acg} suggests that if 
$3p>\mu$  the metric potentials $g_{ab}$   
propagate faster than the speed of light.   
There is emperical evidence suggesting that 
faster-than-light propagation might happen,
see the introduction \ref{sec:intro}.
Energy conditions are a good guide to the physical plausibility of 
fluids because they are macroscopic and energy conditions seem to 
apply in the macroscopic domain;  
also perfect fluids (equation \ref{eq:fluidstress} with $q_{a}=\Pi_{ab}=0$) 
have alot of arbitrariness in the pressure $p$ and density $\mu$ functions 
-  energy conditions narrow the choice of these.  
The microscopic domain is usually described by fields not fluids.   
For the Klein-Gordon field \ref{eq:KG} the weak energy $W$ 
and Ricci convergent energy $C$ are 
\be
W\equiv W^{a}W^{b}T_{ab}=2(\psi_{a}W^{a})^{2}+\psi_{a}\psi^{a}+m^{2}\psi^{2},
\label{eq:Wab}
\ee
and
\be
C\equiv W^{a}W^{b}R_{ab}=2(\psi_{a}W^{a})^{2}-m^{2}\psi^{2},
\label{eq:Cab}
\ee
respectively,   so that the weak energy $W$ increases 
and the Ricci convergent energy $C$ decreases in the presence of a mass term.
The weak energy is a better measure of the 
energy of a field so that it appears that the introduction of mass increases
a fields energy.   
Now the question arises:  
Does this happen for the introduction of non-metric mass?   
If the weak energy condition is taken to apply as 
in \ref{eq:wec},  then \ref{eq:wec} 
and the timelike convergence condition are taken to apply to $\ac{R}$ 
then no information can be extracted to answer the question;  
futhermore if $P_{ab}W^{a}W^{b}$ is 
constructed then there are several terms of undetermined sign.  The initial 
value problem in the next section suggests that the best dynamics for the 
rescaled theory consist of the field equation \ref{eq:feq},  
the non-metric mass equation \ref{eq:nmass},  
and the trace equation \ref{eq:mtr}.
\section{Mass-like Properties of the Non-metric mass.}
\label{sec:pot}
The mass-like properties of the non-metric mass are investigated by 
comparing both the wave-like linearized solutions of the theory and 
the rate of decay of the object of non-metricity to the wave solutions 
and rate of decay of other massive theories.   The initial value problem in 
the previous section suggests that the best field equations for the 
rescaled theory are the field equations \ref{eq:feq} where the Einstein 
tensor is constructed with a connection which is the sum of the 
Christoffel connection and the semi-metric connections,  
the non-metric mass equation \ref{eq:nmass} 
and the trace equation \ref{eq:mtr}.   
In the presence of a vacuum these equations become the same 
as the vacuum equations of general relativity.

In order to linearize these equations the metric is taken of 
the form of \ref{eq:barm},  but with the background metric being
flat spacetime,
\be
g_{ab}=\eta_{ab}+h_{ab},
\label{eq:etaab}
\ee  
this is the same type of linearization as found in 
Weinberg (1972) \cite{bi:weinberg}.
Substituting \ref{eq:rsm} into \ref{eq:nmass} 
and \ref{eq:expQ} gives 
\be
M+(\fr{1}{2}h_{,a}+\p_{a})Q_{.}^{a}+\ep Q_{a}Q_{.}^{a}=0,
\label{eq:Mexp}
\ee
and
\ber
\ka T_{ab}&=&\fr{1}{2}(\Box^{2}h_{ab}-h^{c}_{.b,ac}-h^{c}_{.a,bc}+h_{,ab}) 
               -\fr{1}{2}\eta_{ab}(\Box h-h^{cd}_{..,cd})\nonumber\\     
          &+&\fr{1}{2}Q_{a,b}-\fr{3}{2}Q_{b,a}
               +\fr{1}{2}Q_{.}^{c}(h_{bc,a}+h_{ac,b}-h_{ab,c})\nonumber\\
               &+&(\fr{1}{2}h_{,c}+\p_{c})Q_{.}^{c}+\fr{1}{4}Q_{c}Q_{.}^{c},
\label{eq:Kat}
\ear
where $h=h^{c}_{.c}$ and the trace equation \ref{eq:mtr} remains unchanged.   
Working in the harmonic gauge 
(equation \ref{eq:PlRy} with $w=\fr{1}{2}$)
and assuming that cross terms in $Q_{a}$   
and $h_{ab}$,  and that square terms in $Q$  are negligible gives
\ber
M&=&-Q^{a}_{.a}=\al T,\nonumber\\
\Box ^{2}h_{ab}&=&-Q_{a,b}+ 3 Q_{b,a}+\eta_{ab}Q^{c}_{.,c}
+2\ka(T_{ab}-\fr{1}{2}T),                
\label{eq:hin}
\ear
To proceed further it is necessary to choose a source.   Choosing a 
Klein-Gordon scalar field as source gives
\ber
M&=&-Q^{a}_{.,a}=\al(-\psi_{c}\psi_{.}^{c}-2\mu^{2}\psi^{2})\nonumber\\
\Box^{2}h_{ab}&=&-Q_{a,b}+3 Q_{b,a}+\eta_{ab}Q^{c}_{.,c}   
    +2\al(\psi_{a}\psi_{b}+\fr{1}{2}\eta_{ab}\mu^{2}\psi^{2}),\nonumber\\
(\Box^{2}-\mu^{2})\psi&=&0.
\label{eq:hcon}
\ear
These have solution
\ber
\psi &=& \exp i(x+\sqrt{1+\mu^{2}}t), \nonumber\\
Q  &=& \fr{1}{4} \exp 2i(x +\sqrt{1+3k\mu^{2}}t),\nonumber\\
M&=& 3k \exp2i(x+\sqrt{1+3k\mu^{2}}t).       
\label{eq:sol}
\ear
Equations \ref{eq:hin} are not consistent for $h_{ab}=0$,  
and the equations for $h_{ab}\ne 0$  are intractable.
The equations \ref{eq:sol} have the properties that:  
the mass is position dependent,  
$Q_{a,b}$ is symmetric i. e. $Q_{a,b}=Q_{b,a}$,  
and $M$ is positive (as implied by the dominant energy condition,  
see paragraph after \ref{eq:mtr}). 
Also $\psi$,  $Q$,  and $M$ 
do not propagate at the speed of light.   
Some alternative linearization schemes are given in the authors 
thesis,  Roberts (1986b) \cite{bi:mdr86b}.

A characteristic of massive field theories is that they decay at a faster
rate than the corresponding massless ones.   
In the simple theory the rate of decay of the fields
is difficult to ascertain because there is no known spherically symmetric 
solution for the stress of the massive Klein-Gordon equation \ref{eq:KG} 
coupled to Einstein's equations \ref{eq:efldeq}.
For the rescaled theory,  it will be shown in section \ref{sec:ss} 
that just the assumption of the 
field equations and the non-metric equation implies the Eddington-
Robertson parameters \cite{bi:weinberg} 
remain unity for any value of the non-metric mass $M$.  
Therefore the rate of decay of the gravitational field does not appear to be 
much effected by the introduction of non-metric mass,   
but it is of interest 
to know about the decay of the object of non-metricity.   
From a fairly general point of view rescale the metric as in
\ref{eq:expQ} $\Omega^{2}=\exp(-Q)$,  
now requiring the metric to be asymptotically flat both before 
and after rescaling implies that $Q\rightarrow\pm 0$  as $r\rightarrow\infty$;
however there is no general way of telling if $Q$ decays as $\fr{1}{r}$ 
or $\fr{exp(-r)}{r}$.
To investigate this it is necessary to resort to exact solutions.  The solution
must be spherically symmmetric with a non-zero trace.   The non-zero trace is 
required so that the trace equation \ref{eq:mtr} 
gives a non-metric mass.  The interior 
Schwarzschild solution has trace $3p-\mu$  
but it is not dependent on the radial 
coordinate and so might give misleading results,  in any case the resulting 
equations are intractable;  therefore the most general static spherically 
symmetric solution to the massless scalar field equations 
(the Klein-Gordon equation \ref{eq:KG} with $V=0$) is used.   

The static spherically symmetric scalar-Einstein solution is
(see Roberts (1985) \cite{bi:mdr85a} and references therein)
\ber
ds^{2}=&-&\exp(-\fr{2\mu}{r})dt^{2}
+\exp(\fr{2\mu}{r})(\fr{\eta}{r})^{4}cosech^{4}(\fr{\eta}{r})dr^{2}\nonumber\\ 
       &+&\exp(\fr{2\mu}{r})cosech^{2}(\fr{\eta}{r})
                  (d^{2}\th+sin^{2}d^{2}{\phi}),\nonumber\\
\phi=&-&\fr{\si}{r},
\label{eq:Wsol}
\ear
where $\eta^{2}=\si^{2}+\mu^{2}$ with $\mu$ is the ADM mass 
and $\si$  the scalar charge.  
The trace equation \ref{eq:mtr} gives
\be
M=\al T = -2\fr{\al\si^{2}}{\eta^{4}}\exp(-2\fr{\mu}{r})sinh^{4}(\fr{\eta}{r})
\label{eq:510}
\ee
For \ref{eq:Wsol},  the $\ep=0$ equation of \ref{eq:nmass} becomes
\be
0=M(x)+\fr{r^{2}}{\eta^{4}}\exp(-\fr{2\mu}{r})sinh^{4}(\fr{\eta}{r})\exp(Q) 
      [(r^{2}Q')'+(\eta-1)Q'].
\label{eq:511}
\ee
Eliminating $M(x)$ from \ref{eq:510} and \ref{eq:511} gives
\be
2\al\si^{2}\fr{\exp(-Q)}{r}=\fr{2Q'}{r}+(\ep-1)Q'+Q".
\label{eq:512}
\ee
which can be put in the form
\be
2\al\si^{2}\fr{\exp(-Q)}{r}=\left(rQ'\exp((\ep-1)Q)\right)',
\label{eq:513}
\ee
this has the general solution
\be
Q=-\fr{k}{r},~~~~~~~~~~~~~~(\ep=1),
\label{eq:514a}
\ee
\be
\exp((\ep-1)Q)=l+(1-\ep)\fr{k}{r},~~~~~~~~~~ (\ep\ne1),
\label{eq:514b}
\ee
where $l$ and $k$ are constants.   This solution can give asymptotically flat 
rescaled spacetimes.   For $\ep\ne\fr{2}{3}$ this differential equation 
has solution
\be
\exp(\fr{Q}{2})=\sqrt{\fr{\al}{2\ep-3}}\fr{\si}{r},
\label{eq:515}
\ee
which is not asymptotically flat as $\exp(-Q)$ 
diverges as $r\rightarrow\infty$.
For $\ep=1$ there are also the solutions
\be
\exp(Q)=3\al\cosh(c-\fr{\si}{\sqrt{3}r}),~~~~~~~~~~~(\al>0),
\label{eq:exp3}
\ee
\be
\exp(Q)= -3\al\sinh(c-\fr{\si}{\sqrt{3}r}),~~~~~~~~~(\al<0),
\label{eq:517}
\ee
where c is a constant.   These solutions are asymptotically flat when
\be
c= arc cosh(\fr{1}{\sqrt{3\al}}),~~~~~~~~~~~~(\al>0),
\label{eq:arcch}
\ee
\be
c= arc sinh(\fr{1}{\sqrt{-3\al}}),~~~~~~~~~~~(\al<0),
\label{eq:arcsh}
\ee
a particular case of the first of these is when $c=0$ and $3\al=1$.   
There are similar solutions with the hyperbolic functions 
replaced by trigonometrical ones.  

The trigonometrical functions would appear to be unphysical because of the 
periodic properties they give to $Q$.  If the dominant energy condition is 
satisfied and the mass is taken to be positive then $Q$   
must negative which does not happen for \ref{eq:exp3}.   
For \ref{eq:515} and \ref{eq:exp3} 
with the conformal factor $\exp(-Q)$
the solution is not asymptotically flat and $Q$ diverges as 
$r\rightarrow\infty$.
This could be overcome by choosing $c$ in \ref{eq:517} 
to be larger than any length scale under consideration.   
Perhaps a better alternative is to relax the
requirement that the dominant energy condition is satisfied 
and choose solution \ref{eq:exp3},  doing this $Q$ is well-behaved.   
After some calculation the Eddington-Robertson parameters for the 
spacetime with conformal factor
\ref{eq:exp3} and with constant $c=0$ are found to be $\al=\ga=1$,  
and $\bt=1-\fr{\si^{2}}{6}$.   $\bt$ is unity in both Schwarzschild 
and for the solution \ref{eq:Wsol},  the smaller $\bt$ has the 
interpretation that the centralfugal force is larger,  thus the solution acts
as though it has more mass than it does.   This observation  
suggests that non-metric theories and conformal factors might have application
to the "missing mass" problem.   In passing it is worth noting that 
Bekenstein's (1974) \cite{bi:bekenstein} trick for producing massless 
conformal invariant scalar field solutions from massless scalar solutions,  
introduces hyperbolic conformal factors similar to the above.   
Using this the whole of the above discussion 
could be carried out for conformal scalar fields in a similar manner.

From Roberts (1991c) \cite{bi:mdr96} equation A3.5 the ADM mass is
\ber
A.^{t(S)}&=&\lim_{r \rightarrow \infty} 
             \exp(-Q)[A.^{t(E)}-\fr{Q'}{r^{2}}g^{(E)}_{\th\th}], \nonumber\\
&=&\mu-\lim_{r \rightarrow \infty} Q'.
\label{eq:518}
\ear
From \ref{eq:514a},  \ref{eq:514b}, and \ref{eq:515}
\ber
Q'&=&\fr{k}{r},~~~~~~~~~~~~~~~~~~~~~~~~~~~~~(\ep=1),\nonumber\\
Q'&=&\fr{k}{r^{2}}\fr{1}{1+(1-\ep)\fr{k}{r}},~~~~~~~~~(\ep\ne1),\nonumber\\
Q'&\approx&\fr{2\si}{\sqrt{3}r}(c-\fr{\si}{\sqrt{3}r}),
\label{eq:519}
\ear
respectively and it is apparent that in the limit 
$r \rightarrow \infty$, $Q'$ vanishes.   
Therefore the ADM mass of solutions 
\ref{eq:514a} and \ref{eq:514b} is still $\mu$.   
For all the rescaled asymptotically flat solutions found the ADM mass 
remains unaltered after rescaling.
\section{Linearized Waves in DeSitter Spacetime.}
\label{sec:ds}
In this section it is shown that the traceless part 
of the linearized metric perturbations
of DeSitter spacetime obey the Fierz-Pauli equation,  
with mass $m^{2}=\fr{2}{3}\La$.   This implies that weak gravitational 
waves travel slower-than-light in DeSitter spacetime,  and are super luminal
in Anti-DeSitter spacetime.

Start with a weak gravitational field in the form \ref{eq:etaab}.   
If the cross terms in $h_{ab}$ are taken to be neglidgible so that the vacuum 
Einstein equations reduce to
\be
\Box^{2}h_{ab}=0,
\label{eq:Boxhab}
\ee
as there is no mass term in this equation weak gravitational waves in
Minkowski spacetime travel at the speed of light.   
Now a similar analysis in DeSitter spacetime leads to the Fierz-Pauli (1939)
\cite{bi:FP} equation \ref{eq:FPeq}
with $m$ real and weak gravitational waves traveling slower-than-light in
DeSitter (1917) \cite{bi:DS} spacetime,  and with $m$ imaginary 
and weak gravitaional waves travelling faster-than-light 
in Anti-DeSitter spacetime.   

DeSitter and Anti-DeSitter spacetime have Riemann tensor
\be
R_{abcd}=\fr{\La}{3}(g_{ac}g_{bd}-g_{ad}g_{bc}).
\label{eq:RieDS}
\ee
where $\La>0$ for DeSitter spacetime and $\La<0$ for Anti-DeSitter spacetime.
Taking DeSitter and Anti-DeSitter spacetimes to be the background
so that equation \ref{eq:barm} aplies \ref{eq:Rd} becomes
\be
R_{ab}=\bar{R}_{ab} +\fr{\La}{3}(4h_{ab}-\bar{g}_{ab}h) 
                    -\fr{1}{2}\Box^{2}h_{ab}.
\label{eq:RicDS}
\ee
Taking
\be
R_{ab}=\La g_{ab},~~~~~~\bar{R}_{ab}=\bar{\La}\bar{g}_{ab},
\label{eq:Labar}
\ee
equation \ref{eq:RicDS} becomes
\be
(\La-\bar{\La})g_{ab}=\fr{\La}{3}(h_{ab}-\bar{g}_{ab}h)
                     -\fr{1}{2}\Box^{2}h_{ab}.
\label{eq:LaLa}
\ee
Assuming that differences in the perturbed value of the cosmological constant
to that of the background value is sufficiently weak so that
$\bar{\La}=\La$ equation \ref{eq:LaLa} becomes
\be
0=\fr{\La}{3}(h_{ab}-\bar{g}_{ab}h)-\fr{1}{2}\Box^{2}h_{ab}.
\label{eq:Lahab}
\ee
The trace of this equation is
\be
(\Box^{2}+2\La)h=0.
\label{eq:a18}
\ee
Defining $\psi_{ab}$ as the traceless part of $h_{ab}$:
\be
\psi_{ab}\equiv h_{ab}-\fr{1}{4}\bar{g}_{ab}h,
\label{eq:a19}
\ee
from \ref{eq:a18} and \ref{eq:a19} $\psi$ 
obeys the Fierz-Pauli equation \ref{eq:FPeq}
\be
(\Box^{2}-\fr{2}{3}\La)\psi_{ab}=0,
\label{eq:psiab}
\ee
with mass
\be
m^{2}=\fr{2}{3}\La.
\label{eq:MLa}
\ee
\section{Solar System Bounds on Non-metric Mass.}
\label{sec:ss}
In this section the Eddington-Robertson parameters of theories involving the
non-metric mass are discussed.   The Eddington-Robertson parameters are 
what are measured in solar system tests of gravitational theories, see for 
example Weinberg (1972) \cite{bi:weinberg},  
in general relativity they are unity.   In order 
to calculate the Eddington-Robertson parameters for the simple theory it is 
necessary to know how the spherically symmetric solution 
for a massive Klein-Gordon scalar field \ref{eq:KG};   
such a solution is not known,  therefore 
no information on the Eddington-Robertson parameters of the simple theory can 
be given.  For the rescaled theory the trace equation \ref{eq:mtr} gives
the non-metric mass dependent on the trace of the stress;  
because the trace of the stress vanishes,  the solar 
system exterior to the sun will be modeled by the Schwarzschild solution as 
in general relativity.   In this section it is shown,  using only the field 
equations \ref{eq:feq}
and the non-metric mass equation \ref{eq:nmass}
(and {\bf not} the trace equation \ref{eq:mtr}),  that 
the non-metric mass can have any value 
while the Eddington-Robertson parameters remain unity.
The Schwarzschild solution in isotropic coordinates is
\ber
ds^{2}=&-&\left(\fr{1-\fr{M}{2\bar{r}}}{1+\fr{M}{2\bar{r}}}\right)^{2}dt^{2}
\nonumber\\
       &+&(1+\fr{M}{2\bar{r}})d\si^{2},
\label{eq:Schiso}
\ear
where
\be
d\si^{2}=d\bar{r}^{2}+\bar{r}^{2}d\th^{2}+\bar{r}^{2}sin^{2}(\th)d\phi^{2}.
\label{eq:iso3}
\ee
To introduce the effects of non-metric mass choose a conformal factor
\be
\Om^{2}=\exp(a\bar{r}^{n}),        
\label{eq:Om2}
\ee
expanding the exponential gives
\ber
ds^{2}=&-&(1+a\bar{r}^{n}+\fr{1}{2}a^{2}\bar{r}^{2n}
                         -2\fr{M}{\bar{r}}\nonumber\\ 
       &-&2Ma\bar{r}^{n-1}-Ma\bar{r}^{2n-1}
                         +2\fr{M^{2}}{\bar{r}})dt^{2}\nonumber\\
       &+&(1-2\fr{M}{\bar{r}}+a\bar{r}+aM\bar{r}^{n-1})d\si^{2}. 
\label{eq:expand}
\ear
Now if $n\ge-3$ the Eddington-Robertson parameters are all unity and 
the solution is indistinguishable from the Schwarzschild solution.  
From \ref{eq:nmass} and  \ref{eq:Om2} the non-metric mass is
\ber
M&=&-\td{\na}_{a}Q_{.}^{a}-\ep Q_{a}Q_{.}^{a}\nonumber\\
 &=&-a\bar{r}^{n-2}(1+\fr{M}{2\bar{r}})^{-5}(1-\fr{M}{2\bar{r}})^{-1}  
  (1-n+\fr{M(n-3)}{4\bar{r}})\nonumber\\
 &-&\ep a^{2}n^{2}\bar{r}^{2n-1}(1+\fr{M}{2\bar{r}}).
\label{eq:atdna}
\ear
For $\bar{r}\ge M$
\be
M\approx - a n \bar{r}^{n-2}\left(1-n+\ep a n \bar{r}^{n}\right).
\label{eq:anbar}
\ee
Now for $n=-3$ and at $r$ a fixed value of $r$ the non-metric mass is
\be
M=3a\left(4-\fr{3a\ep}{\bar{r}_{o}^{2}}\right),
\label{eq:3aep}
\ee
the constant $a$ is arbitrary 
and hence the non-metric mass can be made as large 
as required by choosing a suitable value of $a$.
\section{The Initial Value Problem for the Rescaled Theory}
\label{sec:ivp}
In this section the initial value problem is discussed following 
the treatment in Alder et al (1965) \cite{bi:ABS} for general relativity.   
The conventions used here are Latin indices $a,b,\ldots$  
take the values $0$ to $3$ 
and Greek indices $\al,\bt,\ldots$ take values $1$ to $3$,  
which is the opposite convention from Alder et al (1965) \cite{bi:ABS}.

Considerations are restricted to the rescaled theory with a massive 
scalar field as source.  The field equations \ref{eq:feq},  
the non-metric equation \ref{eq:nmass},  
and the trace equation \ref{eq:mtr} become
\be
R_{ab}=2\psi_{a}\psi_{b}+g_{ab}\mu^{2}\psi^{2},     
\label{eq:a31}
\ee
\be
M=-\td{\na}_{a}Q_{.}^{a}-\ep Q_{a}Q_{.}^{a},
\label{eq:a32}
\ee
\be
M=-\al(\psi_{a}\psi_{.}^{a}+2\mu^{2}\psi^{2}),                    
\label{eq:a33}
\ee
respectively.  Initial data is prescribed on a surface $S$ described 
by the equation $x^{0}=0$.  For the metric initial data
\be
g_{ab},~~~~~g_{ab,o},
\label{eq:ogab}
\ee
are given;  the derivatives of the metric interior to the surface $g_{ab,i}$
can be calculated by differentiation in the surface,  but $g_{ab,00}$    
has to be determined.

A first choice of additional initial data is
\be
Q,~~  M,~~\psi,
\label{eq:Qpsi}
\ee
the derivatives of the object of non-metricity interior to the surface 
$Q$ are given by differentiation in the surface.   The value of $M$ is 
given by the initial data,  then $\psi_{i}$ 
can be calculated by differentiating in the surface,  
and $\psi_{0}$ determined by equation \ref{eq:a33}.   
From \ref{eq:rie} and \ref{eq:ric}
$\ac{R_{ab}}=\td{R_{ab}}+P_{ab}$ where $P_{ab}$ is given by 
\ref{eq:pic}.   $P_{ab}$ is given in terms of $Q$  
and its derivatives which can be determined by the procedure above.   
Thus equation \ref{eq:a31} becomes
\be
\td{R}_{ab}=\fr{1}{2}g_{..}^{cd}(-g_{ad,bc}-g_{bc,ad}+g_{ab,cd}+g_{cd,ab}) 
            +A_{ab}.
\label{eq:Aab}
\ee
Now $A_{ab}$ consists of known data of the metric $g_{ab}$  
and it's first derivatives,
the object of non-metricity $Q$ and its derivatives,  
the scalar field and its derivatives.   
Hence the initial value problem has been reduced to the
same form as that for general relativity as discussed in 
Alder et al (1965) \cite{bi:ABS},   
and thus equations \ref{eq:a31},  \ref{eq:a32},  
and \ref{eq:a33} give a fully determined initial value problem.
                                                                      
A second choice of additional data
\be
Q,~~  Q_{0},~~    M,~~ \psi.
\label{eq:2ndch}
\ee
in the rescaled theory the object of non-metricity is a gradient 
vector,  because the components of the object of non-metricity $Q_{0}$  
can be calculated by differentiating $Q$ in the surface the initial 
data $Q$ is equivalent to the initial data $Q$ and $Q_{0}$.  
Thus this \ref{eq:2ndch} set of additional data is equivalent to 
the first set \ref{eq:Qpsi}.

A third choice of initial data is
\be
Q_{a},~~   \psi,~~   \psi_{0}.
\label{eq:3rdch}
\ee
Then equation \ref{eq:a33} gives $M$ immediately,  
next equation \ref{eq:a32} 
gives $Q_{00}$ and then the problem reduces to that of general relativity 
as above.   

A fourth choice of initial data is
\be
Q_{a},~~  M,~~   \psi,~~  \psi_{0}.
\label{eq:4thch}
\ee
and with this choice it is not necessary to introduce the trace 
equation in the form \ref{eq:a33}.

The shock waves or characteristics of the metric $g_{ab}$ and the 
scalar field are null,  this follows in a straightforward manner 
using the standard approach Synge (1960) \cite{bi:synge} V.7 .   
In the simple and rescaled theory $Q_{a}$  
is a gradient vector and the standard approach works;  
if the object of non-metricity $Q_{a}$ is not a gradient vector,  
but simply a vector obeying a first order equation \ref{eq:Qnm} then a 
shock solution has to be redefined as a solution that does not 
give a unique value for $Q_{a,0}$,  again however the shock solutions 
are null.
\section{The Non-recoverability of the Fierz-Pauli Equation.}
\label{sec:fpeq}
The dynamics of both the simple and the rescaled theories have the unusual
feature of position dependent mass.   This feature is not unique to the 
theories discussed here,  it also occurs in conformally invariant mass 
theories of Page (1936) \cite{bi:page},  Fulton et al (1962) \cite{bi:FRW}, 
and Barut and Haugen (1972) \cite{bi:BH},
and in the fluid symmetry breaking theories of the author,  
Roberts (1989) \cite{bi:mdr89b}.
Except for spins $0$,  $\fr{1}{2}$,  
and $1$ fields the presence of a constant mass
term and covariance seem to be exclusive,  thus to have higher spin fields with
constant mass covariance must be broken in some way.

The idea of covariance breaking is best illustrated by considering weak 
field gravity.   Consider the weak field equations of general relativity in the
harmonic gauge
\be
\Box^{2}h_{ab}=2\ka(T_{ab}-\fr{1}{2}\eta_{ab}T),
\label{eq:a41}
\ee
and the Fierz-Pauli (1939) \cite{bi:FP} equation \ref{eq:FPeq} 
for a massive spin $2$ field.   
Now the weak field equations \ref{eq:a41} are covariant, 
the Fierz-Pauli equation \ref{eq:FPeq} is not,  
and futhermore it cannot be derived 
from a covariant Lagrangian,  see for example Freund et al (1969) \cite{bi:FMS}
or Datta and Rana (1979) \cite{bi:DR}.   
To produce the Fierz-Pauli equation from the weak field 
equations it is necessary to equate 
$2\ka(T_{ab}-\fr{1}{2}\eta_{ab}T)$ to give $m^{2}h_{ab}$.
This is here called ``breaking covariance''.   
There appears to be no neat way of doing this for 
stresses such as those of a perfect fluid or a scalar field;  in this
case covariance breaking appears to be unnatural.   
Observations show that the Fierz-Pauli equation is 
incorrect no matter how small $m^{2}$ in \ref{eq:FPeq} is,  see for 
example Datta and Rana (1979) \cite{bi:DR}.  
The solar system,  exterior to the sun,  
is not a vacuum and there must be a small stress present
(see for example Roberts (1998) \cite{bi:mdr98}): thus in this 
case the observational evidence shows that this small stress does not 
reduce to (or covariantly break into) $m^{2}h_{ab}$ term of the Fierz-Pauli
equation \ref{eq:FPeq}.

In the fluid symmetry breaking theory of Roberts (1989) \cite{bi:mdr89b}
the mass of a vector field $X$ is given by
\be
X=-a\left(-1+\sqrt{1-\fr{1}{3b\mu_{1}}}\right),
\label{eq:a43}
\ee
where $a$ and $b$ are positive real constants,  
and $\mu_{1}$ is a fluid density.   Now the
fluid density is position dependent,  hence the vector fields mass is also 
position dependent.   This is simply overcome by assuming that there is no
appreciable density fluctuation in the region under consideration.   In this 
case there is no covariance breaking because the stress with a constant fluid
density remains covariant.

The above examples motivate the question:  Is it possible to recover the
Fierz-Pauli equation from the non-metric theories discussed here?  
Consider just the $\ep=0$ rescaled theory,  
the other theories being similar,  
re-arranging the dynamical equations gives
\be
2\Box^{2}h_{ab}\we \td{R}_{ab}=\td{\na}_{a}Q_{b}+Q_{a}Q_{b}+\ka T_{ab}  
     +\fr{1}{2}g_{ab}\left(M(1-2\fr{\ka}{\al})-Q_{a}Q_{.}^{a}\right).
\label{eq:dyneq}
\ee
To reduce this equation to the Fierz-Pauli equation \ref{eq:FPeq}  
it is necessary to 
equate the right hand side to $m^{2}h_{ab}$ 
and the same problems that occur doing 
this for weak field general relativity re-occur.  Thus it must be concluded
that there is no natural way of breaking general covariance in order to arrive
at the Fierz-Pauli equation \ref{eq:FPeq}.   
In other words non-metric mass 
theory and massive spin-2 linearized theory are entirely distinct.
The Fierz-Pauli equation \ref{eq:FPeq}
might be derivable from the infrared properties
of other theories,  see Frondsal and Heidenreich (1992) \cite{bi:FH}.
\section{The Cosmological Constant.}
\label{sec:cc}
In this section the extent to which it is possible to re-interpreted the
cosmological constant as due to non-metric mass is investigated.   The 
cosmological constant is predicted to be many orders of magnitude larger than 
the upper bound on its size from cosmological considerations 
Zel'dovich (1968) \cite{bi:zeldovich}. 
The predictions of an anomalous cosmological constant comes from symmetry 
breaking in grand unified theories.   The presence of a cosmological constant 
cannot explain orbital irregularities in the outer solar system,  
Roberts (1987) \cite{bi:mdr87b}.   
Here {\it first} attempts are made to absorb the cosmological constant 
in four dimensions and re-interpreted it as non-metric mass.  
{\it Secondly} attention is turned to higher dimensional theories
(see for example Overduin and Wesson (1997) \cite{bi:OW}),  
where the standard Freund-Rubin \cite{bi:FR} compactification is generalized.

DeSitter spacetime can be expressed in conformally flat form \ref{eq:crs}.
The idea is to absorb the cosmological constant $\La$ by choosing a conformal 
factor $\Om^{2}=\exp(-Q)$ so that the rescaled space has a flat metric and 
non-vanishing object of non-metricity $Q$;   then $Q$ is used to generate 
non-metric mass via equation \ref{eq:nmass}.   

DeSitter spacetime can be put in the explicitly static form and rescaled 
by $\exp(-Q)$ to give
\be
ds^{2}=-\exp(-Q)\left(1-\fr{\La r}{3}\right)dt^{2}
       +\exp(-Q)\left(1-\fr{\La r}{3}\right)^{-1}dr^{2}   
       +\exp(-Q)r^{2}d\Si^{2}.
\label{eq:ssds}
\ee
Defining the luminosity coordinate
\be
R^{2}=\exp(-Q)r^{2},
\label{eq:Qr2}
\ee
gives
\be
ds^{2}=-\exp(-Q)\left(1-\fr{\La R}{3}\right)dt^{2}  
       +(1+\fr{1}{2}RQ')\left(1-\La R \exp(Q)/3\right)^{-1}dR^{2}
       +R^{2}d\Si^{2},
\label{eq:frLaR}
\ee
where ' is $\fr{\p}{\p R}$.  To produce the null form define
\ber
R^{*}&=&\int(1+\fr{1}{2}RQ')(1-\fr{1}{3}\La R \exp(Q))
                     \exp(+\fr{1}{2}Q)dR^{2}\nonumber\\
  &=&\int(R \exp(\fr{1}{2}Q)'(1-\fr{1}{3}\La R \exp(Q))dR,
\label{eq:int1R}
\ear              
note that $R^{*}$ is given by simple arc hyperbolic 
and trigonometrical functions 
depending on the sign of the cosmological constant and the value of $R$.  
Now define $v\equiv t+R^{*},  w\equiv t-R^{*}$ to give
\be
ds^{2}=-\left(\exp(-Q)-\fr{1}{3}\La R\right)dv dw
       +R^{2}(d\th^{2}+\sin(\th)d\phi^{2}).
\label{eq:LaRdv}
\ee

Choosing $\exp(-Q)=1+\fr{1}{3}\La R$ gives a spacetime 
in which the metric is flat;  however there is curvature present,  
the Riemann curvature invariant $\ac{R}_{abcd}\ac{R}_{....}^{abcd}$
is unaltered by the coordinate transformations.   
With the choice $\exp(-Q)=1$ the non-zero curvature is in 
$\td{R}_{abcd}\td{R}_{....}^{abcd}$,  the different choice of $\exp(-Q)$ 
essentially {\it "shifts"} it to $P_{abcd}P_{....}^{abcd}$. 
The non-metric mass equation \ref{eq:nmass} becomes
\be
0=M+\fr{2}{r}Q'\exp(Q)\left(1-\fr{2}{3}\La r\right)
   +Q"\exp(Q)\left(1-\fr{1}{3}\La r\right),
\label{eq:a56}
\ee
where ' is now $\fr{\p}{\p r}$.   
What we would like to happen is that substituting 
$\exp(-Q)=1+\fr{1}{3}\La R$ into \ref{eq:a56} 
we would get an equation for $M$ in terms of $\La$.
If this happened we would be able to say that theories with large cosmological
constant are entirely equivalent to massive theories of gravitation with mass 
of the order of the Planck mass.   Unfortunately this does not happen.  
Substituting gives
\be
0=M-2\La+\fr{2}{3}\La^{2} r^{2},
\label{eq:Lafr23}
\ee
and we have an $r$ dependent mass.  The requirements of both a flat metric and
a space dependent quantity are hard to reconcile.
Ignoring the $r$ dependent terms suggests that non-metric mass is or the
same order as the cosmological constant.

An alternative approach is to first find solutions of the wave equation
\ref{eq:a56}
with constant $M$ and then ask what is the resulting structure of spacetime;  
however the non-metric mass equation \ref{eq:a56} 
has so far proved intractable.  
We could ask whether we get solutions if we require the spacetime 
to be Schwarzschild spacetime.   
This would have the interpretation that the cosmological constant
is equivalent to massive gravitation with 
the Schwarzschild solution around each particle;  
however requiring $\exp(-Q)-\fr{1}{3}\La R=1-\fr{2M}{R}$ produces complex 
equations which again appear to be intractable.

Now the role of non-metric gravitation in higher dimensional theories is 
considered,   in particular its is investigated by generalizing the Freund-
Rubin (1980) \cite{bi:FR} compactification 
c.f. Roberts (1991a) \cite{bi:mdr91a}.   
The bosonic part of the 
Cremmer-Julia-Scherk equations for $d=11$ supergravity are
\be
R_{MN}=(F_{MPQR}F_{N...}^{PQR}-\fr{1}{12}g_{MN}F^{2}),
\label{eq:a58}
\ee
\be
\p_{M}(\sqrt{g^{(11)}}F_{....}^{MN_{1}N_{2}N_{3}}) 
=\ka \ep^{P_{1}P_{2}P_{3}P_{4}Q_{1}Q_{2}Q_{3}Q_{4}N_{1}N_{2}N_{3}}
 F_{P_{1}P_{2}P_{3}P_{4}}F_{Q_{1}Q_{2}Q_{3}Q_{4}},
\label{eq:a58b}
\ee
where
\ber
\ka_{1}&=&\fr{1}{48}\pi G,\nonumber\\ 
a,b,\ldots&=&1,2,3,4,\nonumber\\
M,N,\ldots&=&1,2,3,4,5,6,7,8,9,10,11,\nonumber\\
i,j\ldots&=&5,6,7,8,9,10,11,\nonumber\\
g^{(11)}&=& det g_{MN},\nonumber\\
g^{7}&=&det g_{ij}.
\label{eq:detgij}
\ear
Compactification is considered onto
\be
ds^{2}=d\Si^{2}_{4-}+Z(x_{.}^{a})d\Si_{2}^{2},
\label{eq:Siz}
\ee
where $d\Si^{2}_{4-}$ is the Anti-DeSitter metric,  
$d\Si^{2}_{7}$ is the maximally symmetric 
seven dimensional metric and $Z(x_{.}^{a})$ is a function of $x_{.}^{a}$,  
it is the addition of $Z(x_{.}^{a})$
that generalizes the Freund-Rubin compactification,
$Z=1$ gives the Freund-Rubin solution.  
As in the Freund-Rubin compactification only one
component of $F_{....}^{MNPQ}$ is considered,  namely $F_{....}^{abcd}$.    
Integrating the second equation of \ref{eq:a58b} gives
\be
F_{....}^{abcd}=f(\neq x_{.}^{a}) \ep_{....}^{abcd}(g^{(11)})^{-\fr{1}{2}},
\label{eq:Fabcd}
\ee
absorbing $g^{(4)}$ into $f(\neq x_{.}^{a})$ gives
\ber
F_{abcd}F_{....}^{abcd}&=&f^{2}(\ne x_{.}^{a})
   \ep_{abcd}\ep_{....}^{abcd}|g^{(11)}|^{-1}\nonumber\\
&=&g^{4}|g^{(11)}|^{-1}f^{2}(\neq x_{.}^{a})|g^{7}|\nonumber\\
&=&-f^{2}.
\label{eq:nexa}
\ear
The equation of \ref{eq:a58} now becomes
\be
R_{ab}=-\fr{1}{6}\ka_{1}f^{2}g_{ab},~~~~~
R_{ij}=+\fr{1}{12}\ka_{1}f^{2}g_{ij}.
\label{eq:kafg}
\ee
The first equation of \ref{eq:kafg} has as a solution Anti-DeSitter 
spacetime with cosmological constant 
$\La=-\fr{1}{2}\ka_{1}f^{2}$ dependent on the internal space.

For the internal space to be small it is necessary to have $f^{2}$ large and 
thus a large cosmological constant.   There are two methods of reducing the 
size of the cosmological constant.   The first is to rescale the equation 
\ref{eq:Fabcd} by a factor of $\exp(-Q)$ 
and assume that the resulting small cosmological
constant $\de^{2}$ is independent of the internal indices and defined by
\be
\de^{2} \equiv f^{2} \exp(-Q)
\label{eq:deequiv}
\ee
Substituting this equation into the $11$ dimensional 
non-metric massive wave equation \ref{eq:nmass}
gives
\be
M(x_{.}^{i})=-\fr{2}{f}f_{i.}^{i}-\fr{4}{f}(\ep+3)f_{i}f_{.}^{i},
\label{eq:Mxi}
\ee
and thus the large cosmological constant $-\fr{1}{6}\ka_{1}f^{2}$ 
has been replaced by a small
cosmological constant $-\fr{1}{6}\ka_{1}\de^{2}$ 
with the introduction of the mass \ref{eq:Mxi}. 
The alternative approach is to note that K\"{o}ttler's solution 
(the Schwarzschild solution with cosmological constant) 
is also a solution of the equation 
\ref{eq:Fabcd} and then again attempt to absorb 
the cosmological constant by rescaling the line element.
\section{Cosmology and Massive Gravitation.}
\label{sec:cos}
The simple theory equations are related to Hoyles' (1948) \cite{bi:hoyle}
$C$ field equations which are used in steady state cosmology.   Defining
\be
\fr{1}{2}p=-Q_{c}Q_{.}^{c}-V-3H^{2},
\label{eq:QcQ}
\ee
and
\be
\fr{1}{2}\mu=-Q_{c}Q_{.}^{c}+V+3H^{2},
\label{eq:VH}
\ee
the simple field theory field equations \ref{eq:tG} become
\be
G_{ab}+C_{ab}=\ka T,
\label{eq:Cabka}
\ee
where
\be
C_{ab}=-(p+\mu)U_{a}U_{b}-(p+3H^{2})g_{ab},
\label{eq:muU}
\ee
is Hoyles' $C$ field, $U_{a}$ is a time-like vector,  
$H$ is the Hubble constant,  
$p$ is a pressure,  and $\mu$ is a density.   Hoyle requires that
\be
C_{ab}=\td{\na}_{a}\td{\na}_{b}C,
\label{eq:tdnatd}
\ee
and this is an unusual constraint if applied directly to the 
equations \ref{eq:muU}.

For the rescaled theory note that from the form of the Robertson-Walker
line element
\be
ds^{2}=-dt^{2}+R^{2}(t)\left((1-kr^{2})^{-1}dr^{2}+r^{2}d\th^{2}
                        +r^{2}sin^{2}(\th)d\phi^{2}\right),
\label{eq:sinth}
\ee
is unaltered if the metric is conformally rescaled.   For example if the line
element is multiplied by a conformal factor $\exp(-Q)$ 
and then $dT=\exp(-\fr{1}{2}Q)$ and
$\bar{R}=\exp(-Q)R$ are defined,  the metric remains of the same form.   
Thus massive gravitation does not alter cosmology as the Hubble constant 
becomes defined in terms of $\bar{R}$  instead of $R$ etc.\ldots
Non-metricity has been related to inflation in the work of
Stelmach (1991) \cite{bi:stelmach} and Poberii (1994b) \cite{bi:poberiib}.
\section{Conclusion.}
\label{sec:con}
In section \ref{sec:nmme} 
the approach to massive gravitation advocated here is 
motivated by analogy with the massive wave equations of other field theories.
The resulting theory appears to have nothing in common with the Fierz-Pauli 
equation because:  {\it firstly} it is fully covariant,  
{\it secondly} it is based on the Weyl geometry,  
and {\it thirdly} it does not conflict with observations:
the Fierz-Pauli equation \ref{eq:FPeq} has none of these properties.   
It was shown in section \ref{sec:fpeq}
that there appears to be no natural way of breaking the covariance of the 
non-metric massive theory to produce the Fierz-Pauli equation.

A brief description of semi-metric or Weyl geometries was given in section 
\ref{sec:nmme}.   
The properties of conformal rescaling in this geometry was explained.
Four different equations can be produced by analogy with other massive field
theories,  and it was shown that all four can be expressed by equation 
\ref{eq:nmass}.
The mass in this equation is position dependent,  as is the mass in the 
theories of Page (1939) \cite{bi:page},  Fulton et al (1962) \cite{bi:FRW},  
Barut and Haugen (1972) \cite{bi:BH},  and Roberts (1989) \cite{bi:mdr89b};  
in the first three of these the mass is conformally invariant.   
Two theories which involve the non-metric mass equation 
were suggested.   The {\bf first} is the simple theory,  
which is given by Hilbert's Lagrangian $L_{H}$   
and by the Lagrangian $L_{Q}$ \ref{eq:LQ} which has similar form to the
Klein-Gordon scalar Lagrangian;   
this theory does not have the rescaling property.  
The {\bf second} is the non-metric massive theory of gravitation rescaled
from general relativity.   
It was shown in section \ref{sec:ss} that with the dynamics
given by the field equations \ref{eq:feq}  
and the non-metric equation \ref{eq:nmass} while
the non-metric mass can have any value.  
However it is argued in sections \ref{sec:rescaled} and 
\ref{sec:ivp} that the dynamics are best defined when there 
is an additional equation,  called the trace equation \ref{eq:mtr},  
which equates the non-metric mass to the trace of the stress tensor.   
When the trace of the stress tensor vanishes the non-metric mass vanishes,  
thus in a vacuum the 
field equations are the same as for general relativity.   A consequence of
this is that if the spacetime exterior to the sun is taken to be a vacuum,
then it is still modeled by the Schwarzschild solution.

The major problem with the rescaled theory is to produce a consistent 
set of equations to govern the dynamics.   This problem arises because the 
non-metric equation \ref{eq:nmass} 
can only be derived from a Lagrangian when
$\ep=0$ and the mass term takes the form \ref{eq:mstar}  
- these restrictions are incompatible with the rescaling property.   
The field equations \ref{eq:feq}
and the non-metric equation \ref{eq:nmass} 
consist of $11$ geometrical objects,  $11$ physical quantities,  
and $11$ equations and thus viewed as a linear
problem these equations are sufficient,  however the problem is not linear
and in section \ref{sec:rescaled} additional equations are discussed.   
In most field theories,   for example Maxwell and Proca theories,  
the massive theory produces a trace 
for the stress tensor and the massless does not:  
because of this one of the additional equations \ref{eq:mtr},  
called the trace equation,  which relates the mass
of gravitation to the trace of the stress,  appears to be physically well 
motivated.   The initial value problem set up in section \ref{sec:ivp} 
is most elegantly formulated with the addition of the trace equation.

A difficulty (with the initial assumption that the non-metric equation
should be of the same form as wave equations in other massive theories)  is 
that in other massive theories this is in part motivated by both wave-like
solutions to the wave equation,  
and rates of decay which are faster than in the corresponding massless case:  
it is not immediately clear that the non-metric 
equation \ref{eq:nmass} has the same consequences.   
In section \ref{sec:rescaled} it was remarked
that the trace equation would seem to imply that for a perfect fluid with
$3p>\mu$ the metric potentials propagate faster-than-light.   
In section \ref{sec:ivp}
it was mentioned that the shock wave solutions for all the fields are null,
but even the Klein-Gordon equation in flat space has null shock solutions.
A linearization scheme for the rescaled theory with a massive scalar field,
in which the object of non-metricity,  the mass,  and the massive scalar field
all behave as non-null waves was given in section \ref{sec:pot};  
but to produce an explicit form of the weak field metric proved 
impossible as the equations involved are intractable.    
In section \ref{sec:pot} the rates of decay of the fields
were investigated and the rescaled scalar-Einstein spacetime 
was found to provide a completely tractable model,  
which illustrated that the fields have the desired
properties.   Intuitively it would be expected that the massive theory would
produce a greater inward force than the massless theory of general relativity,
and this might have bearing on the problem of "missing mass" in galactic 
dynamics.   The scalar-Einstein spacetime has all 
the Eddington-Robertson parameters equal to unity;  
however after rescaling the spacetime has the $\bt$ Eddington-
Robertson parameter less than unity.   Using the weak field limit it can be
shown that this implies that the inward force is greater in the massive case.
Thus non-metric massive gravitation allows a qualatitive explanation of the 
missing mass in galactic dynamics.

In section \ref{sec:cc} 
it was shown that it is possible to interpret the theoretically
very high cosmological constant produced in spontaneous compactification as 
due to gravitational mass.   
To do this it was necessary 
to generalize Freund-Rubin \cite{bi:FR} compactification.   
In section \ref{sec:cos} the cosmological consequences of the
theory were discussed.   
For the simple theory,  cosmology can be made in a form
similar to Hoyle's $C$ field cosmology Hoyle (1948) \cite{bi:hoyle}
and Weinberg (1972) \cite{bi:weinberg}.   
For the rescaled theory,  because of 
the conformal invariance of the Robertson-Walker line element,  
the cosmological theory is the same as that for general relativity.
\section{Acknowledgements}
This work has been supported in part by the South African
Foundation for Research and Development (FRD).

\end{document}